\documentclass[11pt]{article}
\usepackage{url}

\usepackage{setspace}
\usepackage{caption} 
\usepackage{mathptmx}
\usepackage[T1]{fontenc}

\usepackage{graphicx}%
\usepackage{hyperref}
\usepackage{multirow}%
\usepackage{amsmath,amssymb,amsfonts}%
\usepackage{amsthm}%
\usepackage{mathrsfs}%
\usepackage[title]{appendix}%
\usepackage{textcomp}%
\usepackage{manyfoot}%
\usepackage{booktabs}%
\usepackage{algorithm}%
\usepackage{algorithmicx}%
\usepackage{algpseudocode}%
\usepackage{listings}%
\usepackage{float}
\usepackage{verbatim} %comments
\restylefloat{figure}
\restylefloat{table}
\usepackage{booktabs}
\usepackage{array}
\usepackage{adjustbox}
\usepackage{lscape}

\usepackage{multirow}
\usepackage{multicol}
\usepackage{hhline}
\usepackage{colortbl}
\usepackage{slashbox}
\usepackage[table]{xcolor}
\DeclareMathSymbol{\mlq}{\mathord}{operators}{``}
\DeclareMathSymbol{\mrq}{\mathord}{operators}{`'}
\usepackage{gensymb}
\newcommand{\mycc}{\cellcolor{gray!20}}
\newcommand{\STAB}[1]{\begin{tabular}{@{}c@{}}#1\end{tabular}}

\usepackage{pifont}
\usepackage{authblk}

\title{Uncovering Key Features for Model-Driven Engineering of Complex Performance Indicators: A Scoping Review}

\author[1]{Benito Giunta\thanks{Emails: \texttt{benito.giunta@unamur.be}, \texttt{corentin.burnay@unamur.be}}}
\author[1]{Corentin Burnay}

\affil[1]{Université de Namur, rue de Bruxelles 61\\
5100 Namur, Belgium\\
NaDI, Namur Digital Institute\\
MInDIT Research Center}

\date{\today}
\begin{document}
\maketitle
% For submission matter
%ORCID iD of Benito Giunta: 0000-0003-3707-6650}
%ORCID iD of Corentin Burnay: 0000-0002-0325-1732}
%ORCID iD of Neil Maiden: 0000-0001-6233-8320}
%ORCID iD of Stéphane Faulkner: }
\begin{abstract}
% Set the context
This paper addresses challenges of designing and managing Complex Performance Indicators (CPI), which amalgamate individual indicators to measure latent, yet crucial business factors like customer satisfaction or sustainability indices. Despite their significant value, designing and managing CPI is intricate; they evolve with rapidly changing business contexts and present comprehension and explanation challenges for end-users. Model-Driven Engineering (MDE) emerges as a potent solution to overcome these hurdles and ensure CPI adoption, though its application to CPI remains an understudied research area. While prior efforts targeted specific CPI modeling objectives, a comprehensive overview of literature advancements is lacking. This study addresses this gap by conducting a scoping review yielding dual outcomes: (1) a comprehensive mapping of modeling features in the literature and (2) a comparative analysis of the coverage offered by the modeling frameworks. These outcomes enhance CPI understanding in academic and practitioner circles and offer insights for future MDE CPI advancements.
\end{abstract}
\section{Introduction}\label{introduction}
%CONTEXT
Businesses strive to define and implement strong strategic plans based on the data at hand, with the ultimate goal to develop competitive advantages and outperform competition \cite{lavalle2011big,akter2016improve}. To achieve this, Business Intelligence (BI) systems have been used to facilitate the gathering and usage of vast quantities of data available internally or externally, and to produce valuable insights for the business decision-makers. The underlying motivation of such BI systems is to improve decision outcome quality by providing managers with proper decision support, helping them to grasp the full complexity of the world and reducing the uncertainty in which their decisions take place \cite{alvarez2005entrepreneurs}. BI is a rich process designing heterogeneous artefacts; this paper specifically focuses on the production of Complex Performance Indicators (CPIs) \cite{joint2008handbook}. A CPI assembles various individual indicators together following a preset model to measure a multidimensional, usually latent variable -- e.g., \textit{customers satisfaction} \cite{zani2013fuzzy} or \textit{corporate sustainability index} \cite{dovcekalova2016composite} -- which cannot be directly measured but can be approximated with a combination of lower-level indicators. CPIs are critical in management sciences \cite{hoffmann2020financial,prendergast1999provision,d2020service,zhang2020reassessing} and offer, among many others, the advantages of summarizing many complementary pieces of data into a single aggregated indicator \cite{raghuramapatruni2017} and explaining complicated business concerns.

Implementing CPIs and visualizations aligned with the actual requirements of decision-makers is critical, yet usually considered to be difficult. The reasons are multiple, among others: data sources are varied and highly heterogeneous \cite{bastidas2021concepts}, the world and business-problems are constantly and rapidly changing \cite{raghuramapatruni2017}, CPIs are inherently intricate and difficult to explain \cite{cherchye2007introduction}, decision-makers have subjective expectations in terms of visualization and interaction with the CPIs \cite{raghuramapatruni2017}, people can feel uncomfortable with data systems \cite{pink2018data} if they do not understand them properly and as a result could prefer to rely on their past experience rather than on the actual data \cite{sanford2010role}. CPI reveals to be more than a ``central tendency-like" measure, more than just a summary of other data. Moreover the CPI construction, one system should also dedicate efforts in managing the CPI not to fall into classical threats of CPI; lack of transparency, poor construction and mathematical soundness, misleading interpretation \cite{joint2008handbook}. To grasp a complex reality, the CPI stands at the intersection of mathematical, technical and management complexities. One system should then support these challenges to properly empower the knowledge and wisdom of the CPI stakeholders. As an answer, \textit{Self-Service Business Intelligence} (SSBI) has been proposed to empower business people and let them produce their own BI outputs with the least any technical knowledge \cite{alpar2016self}. SSBI is promising, but currently offers no support for the self-service production of CPIs. Still, the formalization and specification of CPIs is a cumbersome process and a self-service approach of CPIs is a promising alternative to make it more accessible, with multiple positive side-effects. It would improve transparency \cite{klous2016remove,brous2020trusted} and understanding of end-users \cite{choi2021towards}, who would have control on their CPIs definitions and implementations. It would also reduce lead-time between appearance of a need for a CPI and its implementation \cite{michalczyk2020state}. Finally, users would better grasp where the CPI comes from, increasing their trust in the CPI and therefore the chances of adoption of the CPI \cite{yeoh2010critical}.

%PROBLEM
SSBI is a conceptual approach to BI, shifting the user at the center of the development. However, SSBI itself does not dictate any technical choices or paradigms; one has multiple choices when technically developing a system/artefact under the SSBI lens. One typical way of achieving SSBI is the so-called Model Driven Engineering (MDE) approach \cite{poole2001model,gavsevic2009model,baars2014shaping}. In MDE, users provide to the system (i) a description of what they expect using ``simple-to-understand'' graphical notations and (ii) obtain in return working pieces of software automatically derived from models based on predefined transformations rules \cite{kent2002model}. For our research focus, users would (i) provide their CPI requirements -- what business concepts must be captured by the CPI and how -- under the form of a model/description close to natural language and would (ii) receive the CPI implementation(s) by means of the specific transformation rules. Leveraging MDE for the purpose of CPI raises a number of challenges. The \textbf{first challenge} is that the efficiency of a MDE approach strongly depends on the underlying models they offer to the users \cite{verbruggen2022practitioners}. MDE requires a comprehensive way of documenting CPIs. There exist many different modeling approaches to indicators, but which are scarcely aligned, each proposing its own set of modeling features and relations for specifying CPIs. This could lead to a situation where many models exist with no guarantee that any of them is comprehensive or that they capitalize on the current state of research. The \textbf{second challenge} comes from the fact that users models are central to the MDE. Since the user and the CPIs arise from the business, the approach must ensure a proper mapping of the models with the underlying business intentions. It must be compatible with a vocabulary that is understandable by business people and enable to make links between CPIs and their own business artifacts. Creating alignment between the CPIs solution and the business is essential to generate proper insights for decision-makers \cite{de2008model}. It increases user understanding and eases the CPIs creation; ensuring an appropriate users empowerment. Existing MDE approaches to CPI are very heterogeneous in the way they treat that concern. Besides the Language Expressiveness features (RQ1), and the business alignment (RQ2), an approach must meet the technical pre-requisites of Self-service and its underlying MDE which, if not satisfied, can prevent the proper generation of code, based on the user models. The \textbf{third and last challenge} is that a MDE approach should technically propose artifacts that make possible the uptake from users \cite{bezivin2004search}. The approach should break down the initial complexity and propose MDE features and tooling to grant non-technical end-users with design capabilities and substantial control. 

\subsection{Research Objectives and Questions}
In that context, it is possible to find various frameworks designed for Composite Performance Indicators modelling, and more broadly for indicators modelling. They all hold their own specific research objectives and contexts answering different research gaps. To the best of our knowledge, there is no work sharing a unified mapping of what is undertook by current research. Such mapping would contribute to give access to a structured and unified knowledge base supporting indicator modelling. For that reason, the current work takes the form of a scoping review, aiming to examine a body of literature for a determined topic \cite{arksey2005scoping,munn2018systematic}. Among the 4 types of scoping review proposed by \cite{arksey2005scoping}, the third one better fits our current work whose objective is to explore the findings and coverage of a research areas in order to produce a literature identification and mapping. By tackling an emerging issue, the proposed review is therefore shorter but still benefit from the fresh theoretical reporting and model \cite{webster2002analyzing,arksey2005scoping}. The objective of this review is three-fold.
\begin{enumerate}
    \item The first objective is purely descriptive as required for any review. The current work delineates a body of literature, identify the key resources \cite{klein2020literature}, work and standard methodologies and shed light on the underlying evidences \cite{munn2018systematic} ; 
    \item One  threat of review is to remain purely descriptive (i.e. ``who said what" listing) \cite{klein2020literature,bem1995writing,webster2002analyzing}. To mitigate this threat, the current review fulfills its second objective; producing a structured mapping of the key concepts and characteristics of the identified body of literature. A detailed coverage assessment takes the form of a concept matrix \cite{webster2002analyzing} and is processed for each work identified in the scope of the review. This coverage clarifies concepts, provides definition and structures disparate knowledge to advance theory, inform practitioners, researchers and organisational enablement \cite{arksey2005scoping,klein2020literature} ;
    \item As a direct consequence and third objective, the review proposes the concept matrix as a tool for objective framework for artefact assessment. Artefact can be evaluated and one type of evaluation is called \textit{feature comparison} consisting in checking whether the artefact covers or not a defined list of feature \cite{siau2011evaluation}. One main drawback of this technique, is the subjectivity that lies in defining the checklist of feature \cite{siau2011evaluation}. By using the concept matrix as objective means, the feature comparison is possible and mitigate its risk of subjectivity. In the same perspective, the concept matrix also promote objective benchmarking when comparing two artefacts under the same analysis axis.
\end{enumerate}

The main research question of this scoping review can be described as: ``\textit{What is known from the existing literature about the features that support modeling of Composite Performance Indicators, or more broadly Indicators ?}" This overarching question paves the way of the current work with respect to the research question guidelines of \cite{arksey2005scoping,rai2017avoiding} and can be further refined and explored through the following three sub-questions respectively answering the three MDE challenges described earlier:
\begin{enumerate}
    \item Research Question 1 (RQ1): ``\textit{What are the language constructs/features necessary in a MDE approach to design CPIs properly, according to the literature?}" This question refers to the variety of constructs needed for CPI modeling, later called in this paper as the \textit{Language Expressiveness};
    
    \item Research Question 2 (RQ2): ``\textit{What are the features necessary in a MDE approach to bridge the gap between CPIs and business artifacts of users, according to the literature?}", called later in this work as the \textit{Business User (BU) Empowerment};
    
    \item Research Question 3 (RQ3): ``\textit{What are the technical features required when modeling CPIs to ensure functional MDE components, according to the literature?}", later called in this paper as the \textit{Capabilities and Support for MDE}.
\end{enumerate}

First, this comprehensive view on what is hold by current literature allows a better access and understanding of the relevant knowledge \cite{elhajjar2023automation}. Any consumers, practitioners or scholars, can access through the current work to what is done in terms of modelling indicators features to better comprehend how the literature is shaped. The review provides with the list of the features along with their descriptions that can be used for indicator modelling, new framework definition, or new research directions uncovering. Consumers should therefore no longer dedicate as many resources doing this synthesis work themselves. With the knowledge mapping, the Scoping Review seconds brings deeper understanding of the concepts of a research area to plan for future Systematic Literature Reviews as prescribed by \cite{arksey2005scoping}. Thirdly, the current works presents a comparison of current frameworks which brings sufficient knowledge to understand the coverage and positioning of each framework. Anyone working in the same research direction can benefit from this knowledge to comprehend state of related works or to position their research by working on potential gaps \cite{elhajjar2023automation}.

The remainder of the paper is structured as follows. In Section \ref{SoA}, frameworks are collected from the literature. We first present the collection methodology and then the collection results by describing the resulting frameworks. These frameworks are then in-depth analysed in order to identify the features linked to the research questions and to assess the coverage of each framework over the features. The analysis methodology is presented in Section \ref{FrameworkAnalysis}, and the results of the overall study are presented in Section \ref{FAResults}. In Section \ref{discussion}, the implications of the current study are drawn. The work is concluded and future works presented in Section \ref{conclusion}.
% Framework Detail Table
\begin{figure}[b]
    \centering
    \includegraphics[scale = 0.40]{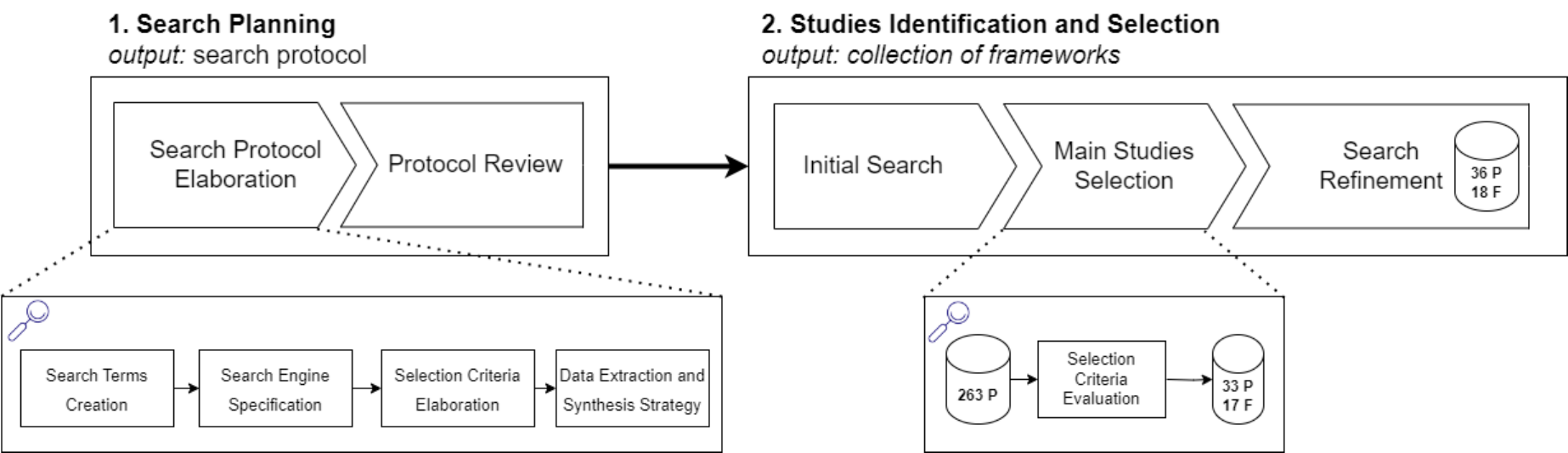}
    \caption{Collection Methodology Process}
    \label{fig:CollectionMethodoProcess}
\end{figure}
\section{Papers Extraction}\label{SoA}
The methodology employed in this review adheres to the guidelines outlined by \cite{arksey2005scoping}. For additional insights, particularly concerning the assurance of study rigor, we turned to the recommendations provided by \cite{kitchenham2004main}. The review unfolds in a structured sequence, commencing with Papers Extraction, followed by Frameworks Analysis, and concluding with the Review Reporting, which involves the documentation and composition of the present work.

The methodology for the papers extraction is depicted in Figure\ref{fig:CollectionMethodoProcess}. Our overall research strategy has consisted in the identification of all the papers advancing a current or possible future MDE approach to CPI, the review of their respective meta-model and the combinations of their results to identify the literature modeling features. Stated differently, we integrate all existing contributions on the topic, with the hope to provide a broader and more systematic view on what MDE CPIs can do, with the underlying assumption that if it was published at least once, it is relevant to consider. All the steps we applied to identify those papers are discussed in the following.

\subsection{Extraction Methodology}
The first step was the \textbf{Search Planning} during which we prepared the collection process, ensuring rigorous and replicable results. First, the \textit{Search Protocol Elaboration} was necessary to define a proper search protocol for the collection of the papers. It divides into four sub-components; (i) we defined the Search Terms Creation to be used for the paper extraction process. We started from our research questions to create three batches of keywords, each batch gathering keywords for its own theme:
\begin{itemize}
    \item One for the concept studied by the framework. We first only searched for ``Complex Performance Indicators" frameworks, but only a few references are specifically dedicated to CPIs (see results later on). Therefore, we extended this batch to incorporate more generic concepts (e.g.: measure, metric, indicator, Key Performance Indicator, etc). The rationale is that a CPI \textit{is an} indicator. The ability of designing an indicator itself is not sufficient but contributes to the design of a CPI; 
    \item One for the field in which the concept is studied (e.g.: Information System, Performance Management, etc); 
    \item One for the activity performed on the studied concept (e.g.: modeling, design, computation, etc).
\end{itemize}
 By combining these groups, we made up a list of search strings broad enough to guide our paper extraction. The list of all keywords is accessible here: \url{https://figshare.com/articles/figure/_/21995219}. (ii) We then decided on the Search Engine Specification; we opted for a search in five main scientific search engines: Google Scholar, IEEE, ScienceDirect, ACM and Springer. (iii) Thirdly, we elaborated our Selection Criteria, used to sort between relevant papers and those not applicable considering our research questions. This resulted in four Selection Criteria conjunctively applied and presented in Table \ref{table:SC}. (iv) Finally, we defined our Data Extraction and Synthesis Strategy to prepare all the model format document to be filled-in later. Once our search protocol was elaborated, we proceeded with a \textit{Protocol Review}, during which we shared our protocol with other researchers from and outside the study to refine and validate it, following the pre-study logic as recommended in \cite{kitchenham2004main}.
 
\begin{table}
\resizebox{\textwidth}{!}{
\begin{tabular}{c>{\arraybackslash}p{13cm}}
\toprule
\textbf{Criterion} & \textbf{Description} \\ 
\hline																
\textbf{SC1}    &	\textit{``The paper contains graphical notation intending to either formalize the indicator constructs, to model the indicator operationalization or to document the indicator.''}\\
\hline
\textbf{SC2}	    &	\textit{``The goal(s) of the framework is explicitly on indicators and serves interests that directly benefit to the definition, management or implementation of indicators.''}\\
\hline
\textbf{SC3}	    &	\textit{``The framework is not domain-dependent, or at least proposes a formalization which is easily transposable to other domains.''}\\
\hline
\textbf{SC4}	    &	\textit{``It is not sufficient that the paper contributes to the literature of indicators and mentions or refers to a certain framework, it should also describe the framework.''}\\
\bottomrule
\end{tabular} 
\caption{Selection Criteria Description}
\label{table:SC}}
\end{table}

The second step of our research was the \textbf{Studies Identification and Selection} during which we applied the search protocol described above. The \textit{Initial Search} aimed to identify a first broad set of papers. For that purpose, we used our different search strings in the different selected search engines. To refine the first batch of papers, we applied our selection criteria to identify the most relevant ones; the \textit{Main Studies Selection}. To do so, we read each paper to fill in an extraction form ensuring systematic documentation; collecting the paper metadata, the concept studied, the method used, some notes about the four Selection Criteria (SC) and the relevancy level of the paper in line with our research questions. Finally, we proceeded with the \textit{Search Refinement} using the ``\textit{Forward Snowballing}'' method from \cite{wohlin2014guidelines}. We started the Forward Snowballing with the papers of the results set of the previous step. For each paper, we identified all papers that cite them; this information can be accessed with Google Scholar. The newly identified papers need to be sorted based on our protocol. For each round, we identified candidate papers on which we applied our triage strategy to add new papers to the results set. The round is repeated for the new papers until no new paper is found.  

\subsection{Extraction Results}
The \textbf{Studies Identification and Selection} initially identified and analyzed 263 papers of which 33 papers remained after applying the selection criteria. It is usual to find multiple papers related to the same framework, i.e. a framework can have various parts or maturity level, each presented in a different paper. The \textit{Search Refinement} took the 33 papers as input and searched for related papers. During the first round, 23 candidate papers were uncovered, of which only 3 were new or conformed with our selection criteria. The second round did not discover additional papers, so that the Snowballing procedure stopped \cite{wohlin2014guidelines}.

The final resulting set consists of 36 papers related to 18 different frameworks. The latter cover a time window of 21 years; from 2003 to 2024. Table \ref{tab:ResultsSet} provides all the details from these frameworks. Note that on the one hand, the \textit{Concept} column describes the granularity of the framework according to its authors -- whether it studies the concept of metric, indicator, Key Performance Indicator (KPI), Performance Indicator (PI) or Complex Performance Indicator (CPI). We noticed that not all authors use the same definitions for these concepts. For the sake of clarity, in the remainder of this work we will adopt and stick to the following definitions. A \textit{metric} is a raw data that is directly collected from its source without any manipulation performed. An \textit{Indicator} is built from metrics, manipulated or not, that are used to monitor a business intention. A \textit{Key Performance Indicator} (KPI) is an indicator which measures one of the factors that are critical for the success of the business \cite{parmenter2015key}. A \textit{Complex Performance Indicator} (CPI) is a composition of several indicators, key or not, according to a predefined model that aims to measure a multidimensional aspect of the business \cite{joint2008handbook}. On the other hand, the \textit{Model} column describes the type of model employed with the framework; an ontology, a meta-model, a pre-conceptual schema \cite{rojas2013executable}, or a Domain Specific Language (DSL).

\begin{figure}[h]
    \centering
    \includegraphics[scale = 0.38]{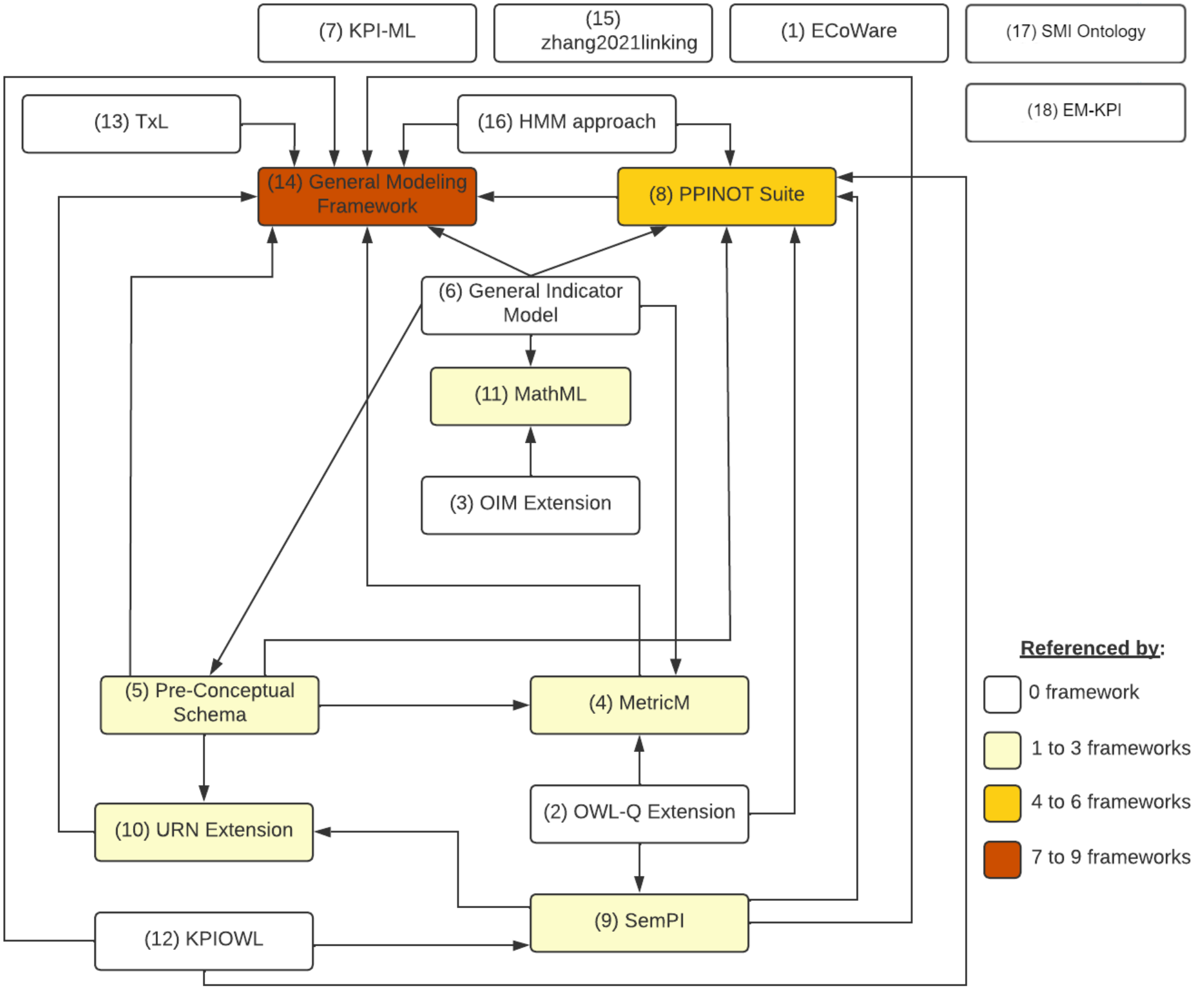}
    \caption{Interrelations within the results set}
    \label{fig:Dependency}
\end{figure}

\begin{table}\centering
\resizebox{1\linewidth}{!}{
\begin{tabular}{p{5cm}|>{\centering\arraybackslash}p{1cm}>{\centering\arraybackslash}p{2cm}>{\centering\arraybackslash}p{2,5cm}}
\hline 
\textbf{ID \& Name} & \textbf{References}  & \textbf{Concept} & \textbf{Model} \\ 
\hline											
\mycc \textbf{(1) ECoWare	}        &	\mycc \cite{baresi2010model}	&	\mycc KPI	&	\mycc ontology\\
\hline									
\textbf{(2) OWL-Q Extension}	&\cite{kritikos2017towards,kritikos2017flexible} &	KPI &	ontology, meta-model	\\
\hline									
\mycc \textbf{(3) OIM Extension}	&	\mycc \cite{letrache2016modeling}	&	\mycc KPI	&	\mycc meta-model	\\
\hline									
\textbf{(4) MetricM	}        &	\cite{frank2009use,frank2008designing,strecker2012metricm}	&	indicators, PI	&	meta-model	\\
\hline									
\mycc \textbf{(5) Executable KPI Pre-Conceptual Schema (KPCS)}	&	\mycc \cite{rojas2013executable}&	\mycc KPI	&	\mycc pre-conceptual schema		\\
\hline									
\textbf{(6) General Indicator Model}	&	\cite{elfouly2015general,elfouly2020flood}	&	KPI, CPI	&	meta-model	\\
\hline									
\mycc \textbf{(7) KPI-ML}	        &	\mycc \cite{brandl2018kpi}	&	\mycc KPI	&	\mycc meta-model	\\
\hline									
\textbf{(8) PPINOT Suite}	&	\cite{del2009towards,del2010defining,del2013definition} \cite{del2013ppinot,del2016using,del2019visual}	&	indicators, PI 	&	ontology, meta-model	\\
\hline									
\mycc \textbf{(9) SemPI}	        &	\mycc \cite{diamantini2014ontology,diamantini2016sempi,diamantini2014collaborative}	&	\mycc KPI, PI &	\mycc ontology, meta-model	\\
\hline									
\textbf{(10) URN Extension}	&	 \cite{pourshahid2009business,pourshahid2011toward,shamsaei2010business}	&	KPI	&	meta-model	\\
\hline									
\mycc \textbf{(11) MathML}	        &	 \mycc \cite{caputo2010kpi}	&	\mycc Indicators, KPI, PI	&	\mycc meta-model	\\
\hline									
\textbf{(12) KPIOWL	}        &	\cite{del2021ontology}	&	metrics, KPI	&	ontology	\\
\hline									
\mycc \textbf{(13) TxL}	            &	\mycc \cite{monahov2013design}	&	\mycc KPI	&	\mycc DSL	\\
\hline									
\textbf{(14) General Modeling Framework (GMF)}	&	 \cite{popova2007specification,popova2010modeling,popova2011formal}	&	indicators, PI		&	meta-model\\
\hline									
\mycc \textbf{(15) KPI Calculation}	&	\mycc \cite{zhang2021linking}	&	\mycc KPI	&	\mycc ontology	\\
\hline									
\textbf{(16) HMM Approach}	&	\cite{van2017transforming}&	PI	&	meta-model	\\
\hline
\mycc \textbf{(17) SMI Ontology}	& \mycc	\cite{de2003towards}& \mycc	indicators	& \mycc	ontology	\\
\hline
\textbf{(18) EM-KPI}	&	\cite{li2019enhancing,ma2024ontology,li2022semantic,li2017identifying}&	KPI	&	ontology	\\
\hline
\end{tabular}
\caption{Results Set}
\label{tab:ResultsSet}}
\end{table}
All frameworks stand for possible approaches for organizational performance management. Some of them adopt the specific view of \textit{Business Intelligence} (3, 4), others of the \textit{Enterprise Architecture Management} (13). Also in line with the organizational performance, some frameworks focus on \textit{Business Processes} (2, 8, 10, 16), on \textit{Services} (1, 2), or on issues related to \textit{Shared Data} (9, 12, 15). Some frameworks also adopt a more specific positioning like the \textit{Smart Manufacturing} (7), the \textit{Logistics} (14), the \textit{Building Performance Management} (15, 18), \textit{web system management} (17) and the \textit{GeoInformation Systems} (6). We keep these frameworks in our set of frameworks because the approaches they advance are still highly generalizable. Even if originating from different fields, it is interesting to note that these frameworks are not completely isolated from each other. Figure \ref{fig:Dependency} shows the citation relationships between them; one framework is represented by one rounded rectangle, and an arrow means its originating framework cites the targeting framework. We can observe that most of the frameworks are interrelated and that there are two core frameworks; the (14) and the (8). 
\begin{figure}[h]
    \centering
    \includegraphics[scale = 0.60]{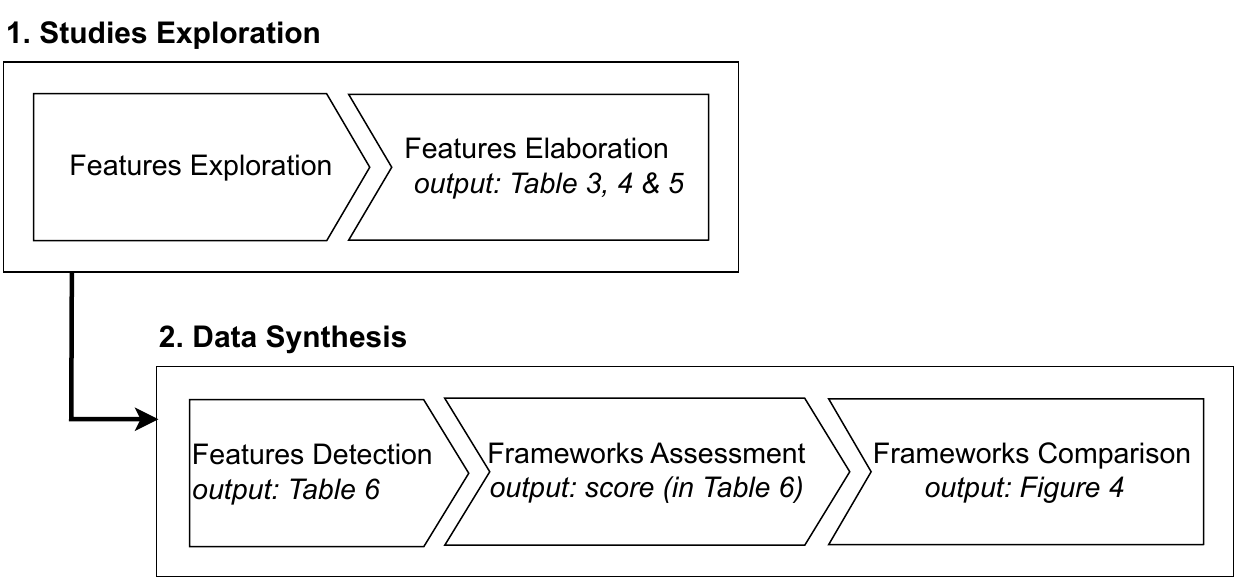}
    \caption{Overall \textit{Analysis Methodology} Process}
    \label{fig:AnalysisProcess}
\end{figure}
\section{Frameworks Analysis Methodology}\label{FrameworkAnalysis}
Based on the selected papers of the previous section, we performed a systematic analysis of the propositions made by each framework regarding the three research questions. This process was strictly guided following the procedure depicted in Figure \ref{fig:AnalysisProcess}, to ensure its replicability and transparency. This section provides details on this analysis protocol, while results of its application are discussed in section \ref{FAResults}.

The initial phase of our study involved the preparation of the analysis, encompassing the development of an analysis protocol and a comprehensive pre-reading of each framework. Subsequently, we commenced our analysis with the Studies Exploration, which entailed the Features Exploration. Here, we systematically identified distinct features characterizing each framework, crucial for subsequent comparisons. The selected frameworks were randomly distributed among the authors, with each author meticulously examining and documenting explicit characteristics contributing to our research questions. Additionally, a concise description of each feature was provided by the researchers.

Following individual assessments, the collaborative phase of Feature Elaboration ensued. Results were amalgamated to create a unified list of features. Features elicited by all researchers underwent discussions to formulate a shared definition, enhancing the robustness of the Studies Exploration. Conversely, features identified by only one researcher prompted thorough discussions until a consensus was reached, aligning with our research questions and contextual relevance.

Based on the retained features, we categorized them into three groups aligned with our research questions: Language Expressiveness, BU Empowerment, and Capabilities and Support for MDE. The objective was to establish a structured dataset for coherent visualization. Group assignments were deliberated among researchers, ensuring agreement on feature allocation. In cases of potential overlap, features were assigned to the group where their contribution was most significant, requiring consensus from all researchers to proceed.

The subsequent step involved Data Synthesis, commencing with Features Detection to evaluate the extent of coverage for each framework concerning the identified features. To enhance objectivity, we synthesized results using a concept matrix, facilitating framework comparisons. Each framework was assessed by researchers, and a Framework-Feature Matrix was created to denote explicit support (\checkmark), absence, or uncertainty (?). Any discrepancies prompted iterative steps, including new feature creation and subsequent reviews by an external BI researcher to refine relevance.

Continuing with Frameworks Assessment, our focus was on evaluating each framework relative to our research questions. A coverage score was computed for each framework and research question group, indicating how well a framework addressed the corresponding research question. Utilizing the Kronecker Delta, we established this score, comparing the variable \(x_{ij}\) representing a framework holding a feature to the true value (1), signifying feature presence in the literature. This function, written \(\delta xy\), returns 1 if \(x_{ij}\) is equal to 1, 0 otherwise.
\[
    score_{x_{j}} =(\frac{\sum_{n=1}^{i}\delta(x_{ij},\mbox{1})}{ \sum_{n=1}^{i}\delta(x_{ij},\mbox{1}) +  \sum_{n=1}^{i}\delta(x_{ij},\mbox{0})}) * 100
\]
\[
    \mbox{\textit{where} } | j \in [1;3]:d_{j} \in \{\mbox{Language Expressiveness, BU Empowerment,}
\]
\[
\mbox{Capabilities and Support for MDE}\}|
\]
\[
    \mbox{\textit{where} } | i \in \{1,...,n\}:x_{ij} = i^{th}  \mbox{ feature of }j^{th}\mbox{ research question}|
\]

As defined by the equation above, for the research question \textit{j}, the score for a framework \textit{x} is the number of features of the research question \textit{j} that the framework holds (when the feature has a V-mark this is accounted for 1), divided by the number of features of the research question \textit{j} that the framework holds plus the number of features of the research question \textit{j} that the framework does not hold (when the cell is empty this is accounted for 0). The result is then converted in a percentage score. The research questions range from 1 to 3 corresponding to \textit{(1) Language Expressiveness}, \textit{(2) BU Empowerment}, \textit{(3) Capabilities and Support for MDE}, and for a certain research question, the features range from \textit{1} to \textit{n}, with \textit{n} the maximum number of features for that specific research question. After all scores are computed, we added them to the previous Framework-Feature Matrix, for all frameworks across all research questions. With this updated matrix, we proceeded with the \textit{Frameworks Comparison}. The aim was to compare the frameworks alongside the three research questions, by means of their scores, to identify the existence of a potential research gap. The visualization to be used should be sufficiently expressive to capture the three research questions, gathering the 16 frameworks under the same figure, without being too complex to still be able to draw conclusions from it. 
\begin{table}[h]
\begin{adjustbox}{center}
\resizebox{1\linewidth}{!}{
\begin{tabular}{c|p{3.5cm}|p{17cm}}
\toprule
&\textbf{Feature} &	\textbf{Description}	\\ 
\midrule
\midrule
\multirow{34}{*}{\STAB{\rotatebox[origin=c]{90}{\textit{Dim. 1:} Language Expressiveness}}}&\textbf{01 -- Indicator Attributes}	&	The constructs of the framework support the adjunction of additional pieces of information to an indicator. E.g.: the name of the indicator, its creation date, etc.	\\ \cmidrule{2-3}
&\textbf{02 -- Traceability}	&	The framework allows direct link between the indicator and the issuing-entity (a department, a business unit, a role, a team, a person,...).  	\\ \cmidrule{2-3}
&\textbf{03 -- Categories of Indicators}	&	The constructs of the model support the existence of sub-categories of indicator. E.g.: The framework (1) specifies a Reliability KPI collecting the portion of correct execution of a service. 	\\ \cmidrule{2-3}
&\textbf{04 -- Operators}&	The constructs of the model formally define the concept of basic arithmetic operators that can be used to specify a metric or an indicator. 	\\ \cmidrule{2-3}
&\textbf{05 -- Formula}	&	The constructs of the model formally define and allow to manipulate the concept of mathematical formula that describes a metric or an indicator.	\\ \cmidrule{2-3}
&\textbf{06 -- OLAP Operations}	&	The constructs of the framework support Online Analytical Processing (OLAP) operations such as basic aggregation operators, drill down, roll-up, etc \cite{vaisman2014}.	\\ \cmidrule{2-3}
&\textbf{07 -- Business-rules}	&	The constructs of the model support the formal definition of conditional constraints that the indicators must comply with.	\\ \cmidrule{2-3}
&\textbf{08 -- Data Source}	&	The framework supports the definition of data sources to be used for the computation of each indicator.	\\ \cmidrule{2-3}
&\textbf{09 -- Value Type and Domain}	&	The constructs of the model allow access and manipulation of the data type of indicators (e.g.: numerical, textual, binary, etc) and/or the range of values the indicators can take within that data type.	\\ \cmidrule{2-3}
&\textbf{10 -- Unit}	&	The constructs of the models allow to associate to and retrieve from an indicator its data unit (e.g. a temperature indicator with its data unit ``$^{\circ}$C'').	\\ \cmidrule{2-3}
&\textbf{11 -- Weights}	&	The constructs of the model allow the use of weighting with indicators.	\\ \cmidrule{2-3}
&\textbf{12 -- Limit Values}	&	The constructs of the model support the definition of lower and/or upper bound limits on the indicators. 	\\ \cmidrule{2-3}
&\textbf{13 -- Status}	&	The constructs of the model allow the definition of a label that communicates the actual state of the indicator. The latter can be expressed either in terms of \textit{Limit Values} (feature 12) that are reached also specifying whether the indicator is a reverse one or not, or the active state of the indicator (e.g.: failed or past, in progress, success, etc). 	\\ \cmidrule{2-3}
&\textbf{14 -- Hierarchy}	&	The framework allows the formal set-up of indicators through hierarchical links, which can be assimilated as a parent-child relationship where the child inherits the features of its parent(s).	\\ \cmidrule{2-3}
&\textbf{15 -- Composition of Indicators}	&	The framework permits formal and dynamic creation of indicators which are a combination of other indicators. 	\\ \cmidrule{2-3}
&\textbf{16 -- Other relationships}	&	The framework formally defines and dynamically handles the creation of links between indicators of any forms other than the parent-child (feature 14) and the composition relationships (feature 15). E.g.: a lagging relationship (framework 4), a dependency relationship (framework 15), etc. 	\\ 
\bottomrule
\end{tabular}}
\end{adjustbox}
\caption{Features Descriptions -- Research Question (1)}
\label{tab:FeatureDescr1}
\end{table}
\section{Frameworks Analysis Results}\label{FAResults}
The first step of the analysis (Figure \ref{fig:AnalysisProcess}) -- the \textbf{Features Creation} -- produced both the \textit{Features Descriptions}, with a total of 31 features. These are all features, found among our collection of frameworks, which share a link with our research questions. In other words, the Features Descriptions encompass and describe all elements that a MDE approach to CPIs should cover, according to the related literature. These features are structured according to our 3 research questions; the Language Expressiveness reported in Table \ref{tab:FeatureDescr1}, the BU Empowerment reported in Table \ref{tab:FeatureDescr2} and the Capabilities and Support for MDE reported in Table \ref{tab:FeatureDescr3}. The analysis also produced a \textit{Framework-Feature Matrix} (Table \ref{tab:RQ1.1}) which synthesizes the coverage of each framework. The matrix shows whether a framework covers a feature (a V-mark), does not cover a feature (a blank cell) or does not share enough information to conclude (a question-mark), for each framework and each feature. The information is structured by the groups of features.

\begin{table}
\begin{adjustbox}{center}
\resizebox{1\linewidth}{!}{
\begin{tabular}{c|p{3,5cm}|p{17cm}}
\toprule
 & \textbf{Feature} 	&  \textbf{Description}	\\ 
\midrule
\midrule
\multirow{22}{*}{\STAB{\rotatebox[origin=c]{90}{\textit{Dim. 2:} Business User Empowerment}}}&\textbf{17 -- Creation of Categories}		&	The framework allows business users to create their own \textit{sub-categories} of indicators (feature 03).	\\\cmidrule{2-3}
&\textbf{18 -- Business Goal Model}	&	The framework somehow does a mapping between the strategic intent \cite{kitsios2018} of the organization under any kind of forms (model, documentation, etc) and the indicators. This feature helps business users to better grasp which business objective is served behind an indicator.	\\\cmidrule{2-3}
&\textbf{19 -- Business Process Model}	&	The framework somehow does a mapping between the business processes under any kind of forms (model, documentation, etc) and the indicator. This feature makes explicit the context -- i.e. the business process -- in which an indicator takes place.	\\\cmidrule{2-3}
&\textbf{20 -- Conceptual Model}		&	The framework somehow does a mapping between the various business entities that are involved in the organisation and the indicators that capture and monitor them.	\\\cmidrule{2-3}
&\textbf{21 -- Organizational Structure Model}	&	The framework formalizes somehow does a mapping between the hierarchical levels present in the enterprise architecture and the indicators. It may take the form of strategic levels or hierarchical roles within the organisation. Both show the business level that is involved in the indicator. 	\\\cmidrule{2-3}
&\textbf{22 -- Maturity}		&	This feature captures whether the framework had been applied through a real-world case study sufficiently advanced to demonstrate the feasibility and the compliance of the framework regarding a real business setting. We assume that, if the framework had been applied in a real-world setting, it certainly is easier for a business user to understand its rationale. We do not claim that the maturity of a framework is limited to the case study application. On the one hand, we consider the inclusion of the maturity aspect as relevant, on the other hand the case study revealed to be the only aspect that serves the maturity.	\\\cmidrule{2-3}
&\textbf{23 -- Design Interface}	&	The framework covers an user-friendly interface which facilitates the exploitation of the artifacts of the \textit{Graphical Notation} (feature 31). It allows non-technical users to design the indicators either in natural language or with any mechanism presented by the authors as especially intelligible by non-technical users. It allows NTU to be empowered in the creation of indicators.	\\
\bottomrule
\end{tabular}}
\end{adjustbox}

\caption{Features Descriptions -- Research Question (2)}
\label{tab:FeatureDescr2}
\end{table}

\begin{figure}[h]
    \centering
    \includegraphics[scale = 0.28]{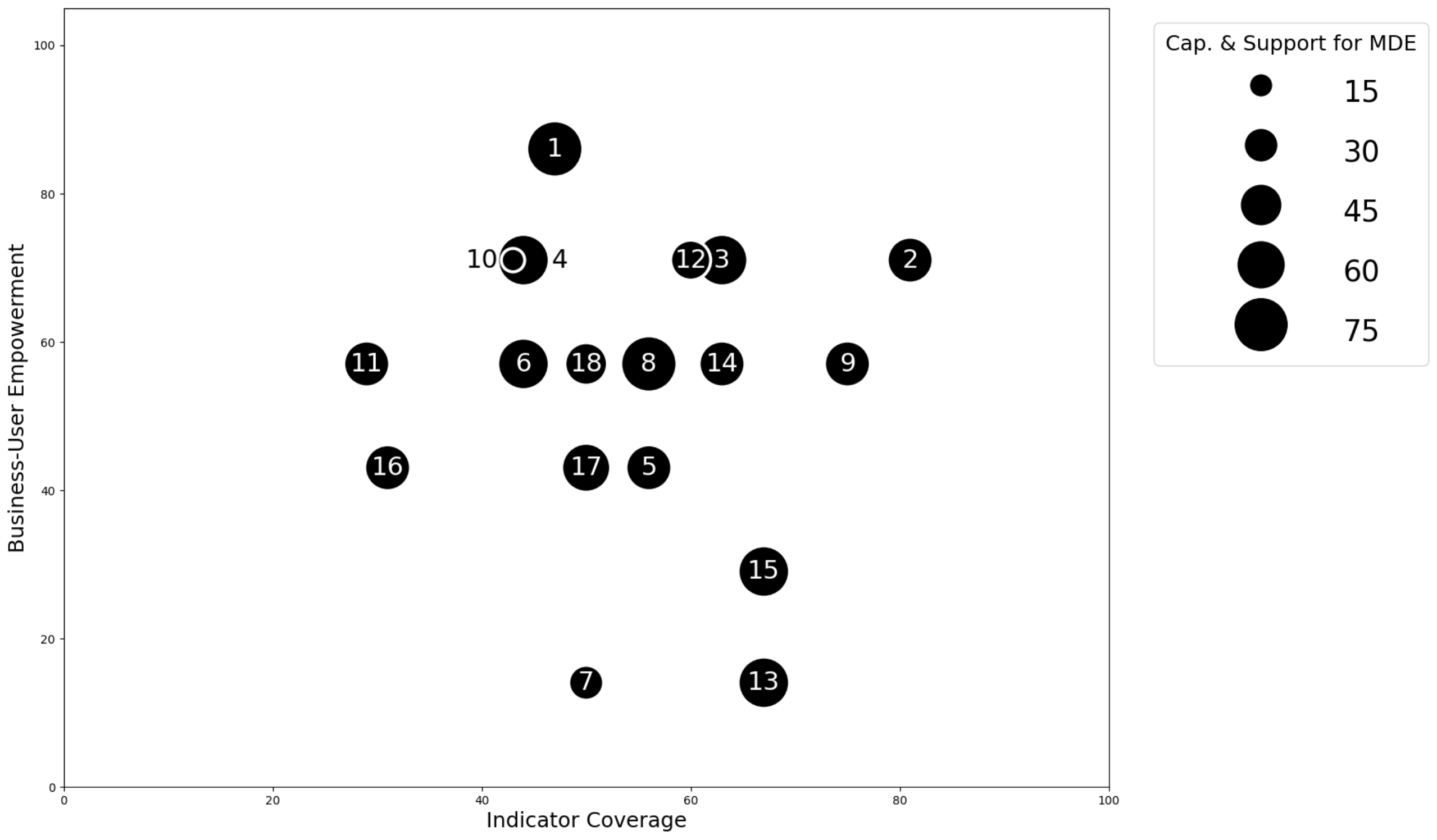}
    \caption{Frameworks Comparison}
    \label{fig:FmkComp}
\end{figure}

\begin{table}
\begin{adjustbox}{center}
\resizebox{1\linewidth}{!}{
\begin{tabular}{c|p{3,5cm}|p{17cm}}
\toprule
 & \textbf{Feature} 	&  \textbf{Description}	\\ 
\midrule
\midrule
\multirow{24}{*}{\STAB{\rotatebox[origin=c]{90}{\textit{Dim. 3:} Capabilities \& Support for MDE}}}&\textbf{24 -- Semantics}	&	The framework permits to specify a semantic layer to indicators involved in its models. It allows dynamic manipulation and reasoning with the indicators making the modeling even more powerful \cite{harel2004,kritikos2017flexible}.\\\cmidrule{2-3}
&\textbf{25 -- KPI Requirements}	&	The framework supports the documentation of requirements for future indicators, be they functional or non-Functional requirements \cite{chung2012}, independently from the model to-be.	\\\cmidrule{2-3}
&\textbf{26 -- KPI Computation}	&	The framework discusses the implementation of the indicators defined by the user. This feature is evaluated to false if, for instance, the framework loads the final value of an indicator, to true if the framework performs the computation of that value.	\\\cmidrule{2-3}
&\textbf{27 -- Runtime}		&	The models are partially or totally executed.	\\\cmidrule{2-3}
&\textbf{28 -- KPI Visualisation}		&	The framework exposes the value of the indicator to the user with functionalities that allow KPI assessment, i.e. tasks that aim to investigate the final value of an indicator instance (e.g.: to check the business conclusion that can be taken from the indicator value).	\\\cmidrule{2-3}
&\textbf{29 -- MD Component}		&	The framework approaches the problem of indicator modeling by adopting explicitly a Model-Driven perspective allowing partial or total generation of code, either with a Model-Driven Engineering (MDE), a Model-Driven Development (MDD), a Model-Driven Architecture (MDA) or a Model-Based Development (MBD) component. It opens the door to the empowerment of non-technical users with a ready-to-use framework.	\\\cmidrule{2-3}
&\textbf{30 -- Query Language and/or Code}	&	The framework comes with workable formalism that supports users (or allows future support) in the specification or interactions with the indicator. It takes the form of either a programming language that implements the indicator, a structured data file (e.g. XML) that stores the definition of the indicator, or a querying language to extract features of the indicator. This feature contributes to increasing the level of non-technical user empowerment.	\\ \cmidrule{2-3}
&\textbf{31 -- Graphical Notation}	&	The framework includes a graphical notation which models either the indicators definition or operationalization. It allows a greater understanding of the indicators by the non-technical users without exposing them to chunks of code or technical stuff. This notation may be the specification of an existing \textit{Design Interface} (feature 23), the foundation for a future design interface or other tooling.	\\
\bottomrule
\end{tabular}}
\end{adjustbox}

\caption{Features Descriptions -- Research Question (3)}
\label{tab:FeatureDescr3}
\end{table}
The second and final step -- the \textbf{Research Questions Analysis} -- produces a \textit{coverage score} for each framework on each research question and a \textit{Frameworks Comparison} scatter plot to compare all frameworks through the three research questions. In Table \ref{tab:RQ1.1}, a score is added for each framework for each group of features. For instance, the framework (1) has the following scores: 47\%, 86\%, 75\% respectively for the \textit{(1) Language Expressiveness}, the \textit{(2) BU Empowerment} and the \textit{(3) Capabilities and Support for MDE}. It means that the framework (1) covers 47\% of the features of the literature regarding the \textit{Language Expressiveness} (RQ1), the same reasoning is applied for the other scores. With the \textit{Frameworks Comparison}, we compare all the frameworks together. A scatter plot is produced for that purpose (Figure \ref{fig:FmkComp}); a circle data point represents a framework, the x- and y-axis respectively represent the score of RQ1 and RQ3 and the size of a data point is proportional to the score of RQ2. Overall, no framework is positioned at 100\% on all axis. The coverage of all frameworks is quite broad with an amplitude of 52, 72, 62 respectively for the x-axis, the data point size and the y-axis. It is interesting to notice that each framework has its own coverage and position which is in most cases unique, except for the frameworks (10, 4) and (12, 3). 
%{\setlength{\hoffset}{-2cm}
%\begin{landscape}
\begin{table*}[h]\centering
%\begin{adjustbox}{max width=\linewidth+2cm, max height=\textheight+2cm} %\textwidth
%\ra{1}
\begin{adjustbox}{center}
\resizebox{1.0\linewidth}{!}{
%\rowcolors{2}{lightgray}{}
\begin{tabular}{r|r|c|c|c|c|c|c|c|c|c|c|c|c|c|c|c|c|c|c}
\toprule
%%%%%%%%%%%%%%%%%%%%%%%%%%%  keep

\multicolumn{2}{c|}{\backslashbox{\textbf{Feature}}{\textbf{Framework}}}	&	\textbf{(1)}	&	\textbf{(2)}	&	\textbf{(3)}	&	\textbf{(4)}	&	\textbf{(5)}	&	\textbf{(6)}	&	\textbf{(7)}	&	\textbf{(8)}	&	\textbf{(9)}	&	\textbf{(10)}	&	\textbf{(11)}	&	\textbf{(12)}	&	\textbf{13}	&	\textbf{(14)}	&	\textbf{(15)}	&	\textbf{(16)}&	\textbf{(17)}&	\textbf{(18)}	\\
\midrule
\midrule

\multirow{16}{*}{\STAB{\rotatebox[origin=c]{90}{Language Expressiveness}}}&01 -- Indicator Attributes	&	\checkmark	&	\checkmark	&	\checkmark	&	\checkmark	&	\checkmark	&	\checkmark	&	\checkmark	&	\checkmark	&	\checkmark	&	\checkmark	&	\checkmark	&	\checkmark	&	\checkmark	&	\checkmark	&	\checkmark	&	& \checkmark & \checkmark	\\

&\mycc 02 -- Traceability        	&	\mycc 	&	\mycc \checkmark 	&	\mycc \checkmark	&	\mycc \checkmark	&	\mycc \checkmark	&	\mycc 	&	\mycc \checkmark	&	\mycc \checkmark	&	\mycc \checkmark	&	\mycc 	&	\mycc 	&	\mycc 	&	\mycc 	&	\mycc \checkmark	&	\mycc 	&	\mycc \checkmark & \mycc\checkmark & \mycc\checkmark	\\
																																	
&03 -- Categories of Indicators	&	\checkmark	&		&		&		&		&		&		&		&	\checkmark	&	\checkmark	&		&		&	\checkmark	&		&	\checkmark	&	\checkmark & \checkmark &	\\

&\mycc 04 -- Operators	&	\mycc 	&	\mycc \checkmark	&	\mycc 	&	\mycc 	&	\mycc 	&	\mycc \checkmark	&	\mycc 	&	\mycc \checkmark	&	\mycc \checkmark	&	\mycc 	&	\mycc \checkmark	&	\mycc \checkmark	&	\mycc \checkmark	&	\mycc 	&	\mycc \checkmark	&	\mycc & \mycc &	\mycc\\																																	
&05 -- Formula	&		&	\checkmark	&	\checkmark	&		&		&	\checkmark	&		&		&	\checkmark	&		&	\checkmark	&	\checkmark	&	\checkmark	&	\checkmark	&	\checkmark	&	& \checkmark & \checkmark	\\

&\mycc 06 -- OLAP Operations	&	\mycc \textbf{?}	&	\mycc \checkmark	&	\mycc \checkmark	&	\mycc 	&	\mycc 	&	\mycc \checkmark	&	\mycc \checkmark	&	\mycc 	&	\mycc \checkmark	&	\mycc \checkmark	&	\mycc 	&	\mycc 	&	\mycc \checkmark	&	\mycc \checkmark	&	\mycc \checkmark	&	\mycc \checkmark & \mycc & \mycc 	\\																																	
&07 -- Business-Rules	&	\checkmark	&	\checkmark	&	\checkmark	&		&		&		&		&		&		&		&		&	\checkmark	&	\checkmark	&	\checkmark	&	\textbf{?}	& & & 		\\

&\mycc 08 -- Data Source	&	\mycc \checkmark	&	\mycc \checkmark	&	\mycc \checkmark	&	\mycc \checkmark	&	\mycc \checkmark	&	\mycc \checkmark	&	\mycc 	&	\mycc \checkmark	&	\mycc \checkmark	&	\mycc \textbf{?}	&	\mycc 	&	\mycc 	&	\mycc \checkmark	&	\mycc 	&	\mycc \checkmark	&	\mycc & \mycc  & \mycc \checkmark 	\\																										
&09 -- Value Type and Domain	&		&	&		&	\checkmark	&	\checkmark	&		&	\checkmark	&	\checkmark	&		&	\checkmark	&		&	\textbf{?}	&	\checkmark	&	\checkmark	&	\checkmark	&	\checkmark	& \checkmark & \checkmark\\																																
&\mycc 10 -- Unit	&	\mycc 	&	\mycc 	&	\mycc 	&	\mycc \checkmark	&	\mycc \checkmark	&	\mycc \checkmark	&	\mycc \checkmark	&	\mycc \checkmark	&	\mycc \checkmark	&	\mycc \checkmark	&	\mycc 	&	\mycc \checkmark	&	\mycc 	&	\mycc \checkmark	&	\mycc \checkmark	&	\mycc \checkmark &  \mycc \checkmark &  \mycc \checkmark	\\																																	
&11 -- Weights	&		&	\checkmark	&	\checkmark	&		&	\checkmark	&		&		&		&	\checkmark	&		&		&		&	\textbf{?}	&		&		&	& & 	\\

&\mycc 12 -- Limit Values	&	\mycc \checkmark	&	\mycc \checkmark	&	\mycc \checkmark	&	\mycc \checkmark	&	\mycc \checkmark	&	\mycc 	&	\mycc 	&	\mycc \checkmark	&	\mycc \checkmark	&	\mycc \checkmark	&	\mycc 	&	\mycc \checkmark	&	\mycc 	&	\mycc \checkmark	&	\mycc 	&	\mycc & \mycc  & \mycc  	\\																																	
&13 -- Status	&	\checkmark	&	\checkmark	&	\checkmark	&		&	\checkmark	&		&	\checkmark	&		&		&	\textbf{?}	&		&	\checkmark	&		&	\checkmark	&		&	& & \checkmark	\\

&\mycc 14 -- Hierarchy	&	\mycc 	&	\mycc \checkmark	&	\mycc \checkmark	&	\mycc 	&	\mycc 	&	\mycc \checkmark	&	\mycc 	&	\mycc 	&	\mycc 	&	\mycc 	&	\mycc \textbf{?}	&	\mycc \checkmark	&	\mycc 	&	\mycc 	&	\mycc 	&	\mycc &  \mycc \checkmark & \mycc   	\\																																
&15 -- Composition of Indicators	&\checkmark	&\checkmark	& &	 	&	 	&	 	&	 \checkmark	&	 \checkmark	&	 \checkmark	&	 	&	 \checkmark	&	 	&	 \checkmark	&	 	&	 \checkmark	& \checkmark & \checkmark & \\

&\mycc 16 -- Other Relationships	&\mycc		&\mycc	\checkmark	&\mycc		&\mycc	\checkmark	&\mycc	\checkmark	&\mycc		&\mycc	\checkmark	&\mycc	\checkmark	&\mycc	\checkmark	&\mycc		&\mycc	\textbf{?}	&\mycc	\checkmark	&\mycc	\checkmark	&\mycc	\checkmark	&\mycc	\checkmark	&\mycc& \mycc  & \mycc 	\checkmark	\\
\bottomrule	

\multicolumn{2}{c|}{\textbf{Research Question (1) -- Score (\%)}}	&	\textbf{47}	&	\textbf{81}	&	\textbf{63}	&	\textbf{44}	&	\textbf{56}	&	\textbf{44}	&	\textbf{50}	&	\textbf{56}	&	\textbf{75}	&	\textbf{43}	&	\textbf{29}	&	\textbf{60}	&	\textbf{67}	&	\textbf{63}	&	\textbf{67}	&	\textbf{31}&\textbf{50}&\textbf{50}	\\
\bottomrule
\multirow{7}{*}{\STAB{\rotatebox[origin=c]{90}{BU Empowerment}}}&17 -- Creation of Category	&	\checkmark	&	\checkmark	&	\checkmark	&	\checkmark	&	\checkmark	&	\checkmark	&		&		&		&	\checkmark	&	\checkmark	&	\checkmark	&		&		&		&	& &	\\																																	
&\mycc 18 -- Business Goal Model	&	\mycc \checkmark	&	\mycc \checkmark	&	\mycc \checkmark	&	\mycc \checkmark	&	\mycc \checkmark	&	\mycc \checkmark	&	\mycc \checkmark	&	\mycc \checkmark	&	\mycc \checkmark	&	\mycc \checkmark	&	\mycc \checkmark	&	\mycc \checkmark	&	\mycc 	&	\mycc \checkmark	&	\mycc 	&	\mycc & \mycc \checkmark & \mycc \checkmark	\\																																	
& 19 -- Business Process Model	&	\checkmark	&	\checkmark	&		&	\checkmark	&		&		&		&	\checkmark	&		&	\checkmark	&		&	\checkmark	&		&	\checkmark	&		&	\checkmark & & \checkmark	\\																																	
&\mycc 20 -- Conceptual Model	&	\mycc \checkmark	&	\mycc 	&	\mycc \checkmark	&	\mycc  &	\mycc 	&	\mycc \checkmark	&	\mycc 	&	\mycc 	&	\mycc 	&	\mycc 	&	\mycc 	&	\mycc 	&	\mycc 	&	\mycc 	&	\mycc 	&	\mycc 	& \mycc \checkmark & \mycc \\																																	
& 21 -- Organizational Structure Model	&	\checkmark	&	\checkmark	&	\checkmark	&	\checkmark	&		&		&		&		&	\checkmark	&	\checkmark	&	\checkmark	&	\checkmark	&		&	\checkmark	&		&	& & \checkmark	\\																																
&\mycc 22 -- Maturity	&	\mycc \checkmark	&	\mycc \checkmark	&	\mycc \checkmark	&	\mycc \checkmark	&	\mycc \checkmark	&	\mycc \checkmark	&	\mycc 	&	\mycc \checkmark	&	\mycc \checkmark	&	\mycc \checkmark	&	\mycc 	&	\mycc \checkmark	&	\mycc \checkmark	&	\mycc \checkmark	&	\mycc \checkmark	&	\mycc \checkmark & \mycc \checkmark & \mycc \checkmark	\\																																	
& 23 -- Design Interface	&		&		&		&		&		&		&		&	\checkmark	&	\checkmark	&		&	\checkmark	&		&		&		&	\checkmark	&	\checkmark	& & \\
\bottomrule																																	
\multicolumn{2}{c|}{\textbf{Research Question (2) -- Score (\%)}}	&	\textbf{86}	&	\textbf{71}	&	\textbf{71}	&	\textbf{71}	&	\textbf{43}	&	\textbf{57}	&	\textbf{14}	&	\textbf{57}	&	\textbf{57}	&	\textbf{71}	&	\textbf{57}	&	\textbf{71}	&	\textbf{14}	&	\textbf{57}	&	\textbf{29}	&	\textbf{43}&\textbf{43}	&\textbf{57}\\
\bottomrule
\multirow{8}{*}{\STAB{\rotatebox[origin=c]{90}{MDE Cap./Support}}}&24 -- Semantics   	&		&	\checkmark	&	\checkmark	&	\checkmark	&	\checkmark	&	\checkmark	&	\textbf{?}	&	\checkmark	&	\checkmark	&		&	\checkmark	&	\checkmark	&	\checkmark	&		&	\checkmark	&	\checkmark & \checkmark &	\\																																	
&\mycc 25 -- KPI Requirements    	&	\mycc 	&	\mycc \checkmark	&	\mycc 	&	\mycc \checkmark	&	\mycc \checkmark	&	\mycc 	&	\mycc \checkmark	&	\mycc 	&	\mycc 	&	\mycc 	&	\mycc 	&	\mycc 	&	\mycc 	&	\mycc \checkmark	&	\mycc 	&	\mycc \checkmark & \mycc \checkmark &	 \mycc \\																																	
& 26 -- KPI Computation	&	\checkmark	&		&	\checkmark	&		&		&		&		&		&		&		&	\checkmark	&		&	\checkmark	&		&	\checkmark	&	& \checkmark & \checkmark	\\																																	
&\mycc 27 -- Runtime	&	\mycc \checkmark	&	\mycc 	&	\mycc \checkmark	&	\mycc 	&	\mycc 	&	\mycc \checkmark	&	\mycc 	&	\mycc \checkmark	&	\mycc \checkmark	&	\mycc 	&	\mycc 	&	\mycc 	&	\mycc \checkmark	&	\mycc 	&	\mycc \checkmark	&	\mycc \checkmark	& \mycc  \textbf{?} &  \mycc \textbf{?}\\																																	
& 28 -- KPI Visualisation	&	\checkmark	&	\checkmark	&		&	\checkmark	&	\checkmark	&	\checkmark	&	\checkmark	&	\checkmark	&	\checkmark	&	\checkmark	&		&	\checkmark	&		&	\checkmark	&	\checkmark	&	& & \checkmark	\\																																	
& \mycc 29 -- MD Component	&	\mycc \checkmark	&	\mycc 	&	\mycc \checkmark	&	\mycc \checkmark	&	\mycc 	&	\mycc \checkmark	&	\mycc 	&	\mycc \checkmark	&	\mycc 	&	\mycc 	&	\mycc \checkmark	&	\mycc 	&	\mycc 	&	\mycc \checkmark	&	\mycc 	&	\mycc 	& \mycc & \mycc\\																																	
& 30 -- Query Language and/or Code	&	\checkmark	&	\checkmark	&	\checkmark	&		&	\checkmark	&	\checkmark	&		&	\checkmark	&	\checkmark	&		&	\checkmark	&	\checkmark	&	\checkmark	&	\checkmark	&		&	\checkmark & \checkmark & \checkmark	\\																																	
&\mycc 31 -- Graphical Notation   &	\mycc \checkmark	&	\mycc 	&	\mycc 	&	\mycc \checkmark	&	\mycc 	&	\mycc 	&	\mycc 	&	\mycc \checkmark	&	\mycc 	&	\mycc 	&	\mycc 	&	\mycc 	&	\mycc \checkmark	&	\mycc 	&	\mycc \checkmark	&	\mycc & \mycc & \mycc 	\\
\bottomrule																																	
\multicolumn{2}{c|}{\textbf{Research Question (3) -- Score (\%)}}	&	\textbf{75}	&	\textbf{50}	&	\textbf{63}	&	\textbf{63}	&	\textbf{50}	&	\textbf{63}	&	\textbf{29}	&	\textbf{75}	&	\textbf{50}	&	\textbf{13}	&	\textbf{50}	&	\textbf{38}	&	\textbf{63}	&	\textbf{50}	&	\textbf{63}	&	\textbf{50} & \textbf{57} & \textbf{43}	\\
%%%%%%%%%%%%%%%%%%%%%%%%%%%  keep
\bottomrule
\end{tabular}}
\end{adjustbox}
\caption{Framework-Feature Matrix}
\label{tab:RQ1.1}
\end{table*}
%\end{landscape}}

\section{Discussion}\label{discussion}
The results of the current work draw a number of implications, discussed hereafter. First of all, the \textbf{Features Descriptions} (Tables \ref{tab:FeatureDescr1}, \ref{tab:FeatureDescr2} and \ref{tab:FeatureDescr3}) translate the three MDE challenges -- the research questions of the current work -- into more operational, functional or non-functional features. It highlights the elements that need to be considered and designed in order to resolve the MDE challenges. Moreover, these features along with their descriptions propose, for each MDE challenges, a response that is the union and summary of the current literature; a unified and rigorous answer to each research question. 

Secondly, the \textbf{Framework-Feature Matrix} (Table \ref{tab:RQ1.1}) and the \textbf{Frameworks Comparison} scatter plot (Figure \ref{fig:FmkComp}) advance several benefits. The first one is offering a view on all features that are shared or not by the frameworks. It raised two types of insights: (i) there are features shared by more than 80\% of the frameworks (feature 01, 18, 22) showing a kind of cross-domain consensus for some features, (ii) there are other features only shared by a minority of frameworks (feature 03, 11, 14, etc) demonstrating the need for an unified view; focusing on only one framework would mean missing relevant features regarding the MDE challenges. The second benefit is the identification of research gaps; the main gap that can be highlighted is the definition of a framework ticking all boxes from the Framework-Feature Matrix (Table \ref{tab:RQ1.1}) and positioned in the upper-right quadrant of the Frameworks Comparison scatter plot (Figure \ref{fig:FmkComp}). Such framework does not exist yet. The third benefit is that the results overall clarify the scientific foundation -- all discovered features -- that a new MDE framework for CPI design should at least cover, which is deeply developed in Section \ref{conclusion}.
Finally, the Framework-Feature Matrix can be leveraged by any interested parties for artefact assessment or comparison through the \textit{feature comparison} technique as mentioned earlier \cite{siau2011evaluation}. The matrix proposes an objective checklist of feature that can be used to assess an artefact that fall into the scope of the current review. For each feature, the artefact can whether support the feature, not support the feature, or not being able to conclude on the feature. 

\section{Limitations}
A number of possible limitations of this paper is worth noticing. First, even if the features are strongly convincing, it may be likely that the current work has not uncovered 100\% of the existing features. This may be induced for two reasons. The first reason, the resulting features strongly depend on the keywords. If one keyword had been added, it might have uncovered additional features. This has been mitigated by the broad scope of the keyword list and the saturation effect identified. The second reason is that some features may exist from a practitioner perspective without being presented in the scientific literature. Such features are not included in the result of this work since the objective of a scoping review is to summarise a body of literature. While empirical features remains highly valuable, an entire study and evaluation need to be undertook to uncover them, please refer to the future works (section \ref{conclusion}). 

Secondly, one may discuss that some features have greater importance/criticality than other. Investigating such information is highly desirable but out of scope of the current work. A scoping review does not intend to quality assessment of the literature evidence. Its main goal is to report in an objective way the state of a research area. Assessing the features importance will be part of the future works (section \ref{conclusion}). 

Thirdly, the result of this work depends on the identification and interpretation of these features -- Table \ref{tab:FeatureDescr1}, \ref{tab:FeatureDescr2} and \ref{tab:FeatureDescr3} -- which had been manually processed. It would be very hard and not necessarily more efficient to do otherwise; an automatic parsing and analysis would definitely be more systematic but would be, on the one hand, very complex and, on the other hand, may reduce the relevancy towards our research focus. To mitigate the potential drawbacks of manual processing, we proceeded in two steps with an adapted setting to ensure a certain level of rigour. As explained earlier, the processing had been randomly distributed across the researchers for both steps individually; meaning that it was not necessarily the researcher who discovered the feature who interpreted it later on, reducing potential biases based on a familiarization effect. Full agreement was always required to proceed, forcing the researchers to raise, fully discuss, and resolve any uncertainty. Finally, a researcher external to the study with scientific Business Intelligence background had proceeded to a complete review with the main task of assessing the features relevancy.

Finally, the use of the Framework-Feature Matrix as tool for \textit{feature comparison} technique \cite{siau2011evaluation} are subject to the limitation of the technique itself; its two-fold subjectivity. The creation of the feature checklist by the author of the artefact to be evaluated is obviously subjective. This threat is reduced as the feature list is now created independently from the artefact creation. However, the analysis - i.e. whether the new artefact covers feature or not - can be subjective. This can be reduced by proposing feature that are clearly defined \cite{siau2011evaluation} and this is where the current work contribute. However, the subjectivity can be further reduce if, for each feature, non ambiguous rules are defined to assess whether the feature is covered, which is left for future works.

\section{Conclusion and Future Works}\label{conclusion}
In this paper, our focus revolves around the complexities associated with CPIs and Model-Driven Engineering (MDE). MDE serves as a Self-Service Business Intelligence approach by increasing end-users involvement while concurrently reducing the complexity faced by them. Numerous approaches exist for designing indicators within the MDE paradigm, each with its unique research objectives. However, there is currently no unified perspective on the ongoing efforts in the literature. Our work aims to fill this gap by presenting a scoping review of relevant frameworks, offering an assessment of the current state of research in this area. We collect, briefly describe, and thoroughly analyze these frameworks. The data synthesis yields three main outputs: (i) Feature Descriptions, collecting, describing, and unifying features from the literature addressing MDE challenges; (ii) a Framework-Feature Matrix, showcasing the coverage of current frameworks; and (iii) Frameworks Comparison, facilitating the assessment of the positioning of each framework.

The results of our study pave the way for future works -- especially, the full elicitation of CPI modeling features in the context of Model-Driven Engineering. The primary objective is to propose a comprehensive set of modeling features guiding users in the CPI modeling process. This set of features should be inclusive, incorporating both literature and empirical features. While the current work provides literature features, a subsequent feature collection with practitioners is necessary to enrich the knowledge base. Additionally, the existence and importance of each feature should be evaluated, possibly through studies involving scholars and practitioners.

If the elicitation of CPI modeling features proves compelling, it sets the stage for a larger-scale future endeavor: the development of a framework for the design of SSBI CPIs leveraging MDE. This framework would incorporate features identified in previous research (both the current work and the future work outlined above) to address the MDE challenges. The new framework would require a language for CPI modeling, supported by a Meta-Model that aligns with the required language constructs uncovered previously. Validation with experts would ensure technical expressiveness, and efforts would be directed towards providing graphical modeling notation to ensure accessibility without a modeling background. Additionally, incorporating features related to user empowerment and MDE capabilities would be crucial, transforming them into consistent functionalities aligned with the CPI language. This research agenda aims to contribute significantly to the advancement of SSBI for CPIs within the MDE context.

% ---- Bibliography ----
%\section*{References}
\bibliography{CPIF_lib}
\end{document}